\documentclass[twocolumn]{aastex7}

\usepackage{siunitx}
\shorttitle{Low-Mass Protocluster Candidate in SMC}
\shortauthors{Meena et al.}

\usepackage{float}
\usepackage{color}
\usepackage{comment}
\usepackage{graphics}
\usepackage{epstopdf}
\usepackage{microtype}
\usepackage{subfigure}
\usepackage[normalem]{ulem}
\usepackage{natbib}
\usepackage{amsmath}

\definecolor{malachite}{rgb}{0.01, 0.8, 0.24}

\usepackage{soul,xcolor}
\setstcolor{red}

\begin{document}
\title{Detection of a Deeply Embedded Protocluster Candidate in NGC 602 with JWST}

\author[0000-0001-8658-2723]{Beena Meena}
\affiliation{Space Telescope Science Institute, 3700 San Martin Drive, Baltimore, MD 21218, USA}
\email[show]{bmeena@stsci.edu}

\author[0000-0002-6091-7924]{Peter Zeidler}
\affiliation{AURA for the European Space Agency (ESA), ESA Office, Space Telescope Science Institute, 3700 San Martin Drive, Baltimore, MD 21218, USA}
\email{zeidler@stsci.edu}

\author[0000-0003-2954-7643]{Elena Sabbi}
\affiliation{Gemini Observatory/NSFs NOIRLab, 950 N. Cherry Avenue, Tucson, AZ 85719, USA}
\affil{Steward Observatory, University of Arizona, Tucson, AZ 85721, USA}
\email{elena.sabbi@noirlab.edu}

\author[0009-0007-8087-6975]{Antonella Nota}
\affiliation{Space Telescope Science Institute, 3700 San Martin Drive, Baltimore, MD 21218, USA}
\affiliation{The International Space Science Institute, Hallerstrasse 6, 3012 Bern, Switzerland}
\email{antonella.nota@issibern.ch}

\author[0000-0003-4196-0617]{Camilla Pacifici}
\affiliation{Space Telescope Science Institute, 3700 San Martin Drive, Baltimore, MD 21218, USA}
\email{camilla.pacifici@gmail.com} 

\author[0000-0003-4870-5547]{Olivia C.\ Jones}
\affiliation{UK Astronomy Technology Centre, Royal Observatory, Blackford Hill, Edinburgh, EH9 3HJ, UK}
\email{olivia.jones@stfc.ac.uk}

\begin{abstract}
JWST NIRCam and MIRI photometry of NGC~602, a low-metallicity young star cluster in the Small Magellanic Cloud, reveals an extended mid-infrared bright emission feature designated as MZS-1. This feature is prominent between 10~$\mu$m and 25.5~$\mu$m, but is extremely faint at 7.7~$\mu$m and entirely undetected at shorter wavelengths. MZS-1 exhibits an elliptical morphology with a major axis of approximately 8\arcsec\ and a minor axis of about~4\arcsec. Its elongated shape and multiple emission peaks in the two-dimensional flux map suggest a group of deeply embedded sources with blackbody-like temperatures ranging from 100~K to 140~K. SED fitting using the \citet{Robitaille2017} model grids identifies these sources as Stage I young stellar objects (YSOs) with masses below~$\sim$3~$M_\odot$ and total stellar mass of the protocluster $\approx$300~$M_\odot$ (based on Salpeter IMF). The low YSO masses are consistent with their absence in Spitzer-based catalogs due to sensitivity limits. By revealing a deeply embedded, low-mass protocluster invisible in previous surveys, this work highlights JWST’s unparalleled resolution and sensitivity in uncovering the earliest stages of low-mass cluster formation in the metal-poor regime.
\end{abstract}


\section{Introduction} \label{sec:intro}

NGC~602~(its H~II region is known as N90) is a young star cluster located near the Magellanic Bridge in the ‘wing’ of the Small Magellanic Cloud (SMC) at a distance of ~61 kpc \citep{Hilditch2005}. It formed in a low-metallicity environment \citep[$\approx1/5~{\rm Z}\odot$,][]{Russell1992, Lee2005} with low gas density \citep[$1.3~{\rm cm}^{-3}$;][]{RomanDuval2014}. Despite these conditions, it exhibits efficient star formation activity and has a stellar mass of $M_\star = 1.6\times10^3~{M}_\odot$ \citep{Cignoni2009}, comparable to young clusters in the Milky Way such as Trumpler-14 in the Carina Nebula \citep{Reiter2019}. The relative isolation of NGC 602, combined with its low dust extinction \citep{Stanimirovic2000}, offers a clear view, making it an ideal laboratory for studying star formation in low-metallicity environments—key for understanding star formation in early-universe like conditions \citep{Kewley2005, Garcia2021} and for developing empirical templates for high-redshift observations.

NGC 602 is thought to have formed roughly 8 Myr ago due to interactions between two expanding H I shells near the boundary between the SMC and the Magellanic Bridge \citep{Nigra2008, Fukui2020}. Optical studies indicate that star formation began around 4–5 Myr ago, with a prominent secondary episode about 2 Myr ago \citep{Carlson2007, Cignoni2009, DeMarchi2013}.
A substantial population of young stellar objects (YSOs) and pre-main-sequence (PMS) stars further confirms ongoing star formation in NGC~602 \citep{Carlson2007, Schmalzl2008, Cignoni2009, Carlson2011, Gouliermis2012}. 

In general, YSOs are classified based on their infrared (IR) spectral index ($\alpha$), which characterizes the slope of their spectral energy distribution (SED) in the IR \citep{Lada1987}. Class I objects ($\alpha \geq 0.3$) are protostars with substantial infalling envelopes; Class II objects ($-2 < \alpha < 0$) correspond to classical T Tauri stars with circumstellar disks; and Class III objects ($\alpha < -2$) represent pre-main-sequence stars with little or no remaining disk material. An even earlier phase of protostellar evolution, termed Class 0, was introduced by \citet{Andre2000}. These Class 0 objects are deeply embedded in dense envelopes, undetectable in the near-IR but observable at mid- to far-IR wavelengths. They exhibit short lifetimes ($<$0.5 Myr) and display a cold, blackbody-like SED dominated by thermal dust emission that peaks in the far-IR to submillimeter (sub-mm) and generally falls below detection thresholds at wavelengths $<$10~$\mu$m.

Due to heavy dust extinction and varying inclinations, misclassifications of young stellar objects (YSOs) based solely on their observed SEDs is common. To address this, \citet{Robitaille2006} proposed a classification based on the physical evolutionary stage of the source rather than its SED shape. In this scheme, Stage I objects (encompassing traditional Class 0/I) are characterized by significant infalling envelopes, Stage II objects have optically thick disks with little to no envelope, and Stage III objects have optically thin disks.

While numerous Stage I and Stage II YSOs have been identified in the LMC and SMC using Spitzer and Herschel \citep{Oliveira2013,Sewilo2013,Seale2014,Ward2017,Oliveira2019}, limited spatial resolution and sensitivity likely caused many deeply embedded or faint sources to be missed, misclassified, or blended with nearby objects.

Using combined HST and Spitzer observations, \citet{Carlson2011} identified over 40 YSOs in NGC 602 spanning evolutionary Stages I to III based on SED modeling. These YSOs are mainly located within dusty ridges of    the H II region. Approximately 70$\%$ of these YSOs have optical counterparts in HST data, but due to Spitzer’s lower resolution (1.6\arcsec–6\arcsec), some sources may be blends, with the optical detections representing more evolved components. Only five sources lack optical signatures, indicating deeply embedded objects; all but one are Stage I YSOs with masses $>$~6~$M_\odot$. All these sources are detected in the near-IR (3–8~$\micron$), and mid-IR at 24~$\micron$.

In this paper, we report the detection of a deeply embedded system in the JWST/MIRI observations of NGC~602. This source likely represents a stage I protocluster in formation—an early-stage structure whose signatures were previously unresolved or undetected in earlier observations. The manuscript is organized as follows: in \S\ref{sec:obs}, we describe the observations used in this study. In \S\ref{subsec:detect}, we present the visual detection of the source, followed by an analysis of its morphology and the possibility of it being a protocluster (\S\ref{subsec:multi}), its observed SED (\S\ref{subsec:sed_fit}) and our efforts to determine its physical properties using YSO modelling (\S\ref{subsec:yso_fit}). We conclude with a discussion of the implications and limitations of this discovery in \S\ref{sec:discussion}.

\section{Observations and Data}\label{sec:obs}
\begin{figure*}[ht!]
\centering
\includegraphics[width=0.95\textwidth]{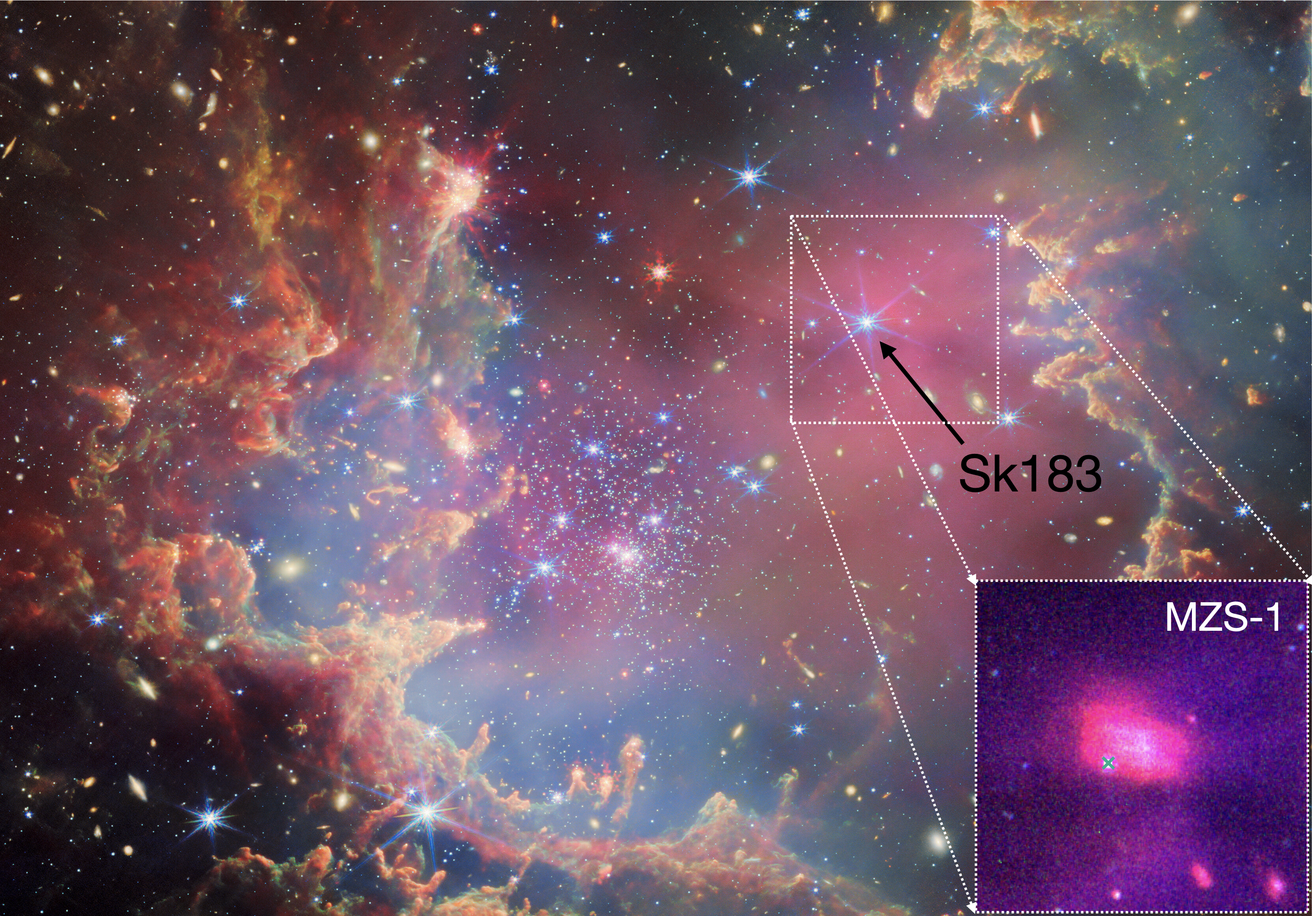}
\caption{A color-composite NIRCam image of the central region of NGC~602, adopted from \citet{Zeidler2024}. The image combines the following filters: F115W+F140M+F150W (blue), F210M (cyan), F277W (green), F335M (yellow), F356W (orange), and F480M (red). The strong mid-IR emission from MZS-1 (discussed in \S\ref{subsec:detect}) is not detected in this NIR image. The bright O3 star Sk183 is indicated for reference. The inset at the bottom right shows an RGB image constructed from MIRI F1000W, F1130W, and F1550W data, highlighting a 25\arcsec$\times$25\arcsec\ region centered on the mid-IR emission, which appears as diffuse pink emission. In this view, Sk183 appears fainter and is marked with a green cross. North is up, and East is to the left. Note that this image does not cover the full NIRCam field of view. Credits: ESA/Webb, NASA \& CSA, P. Zeidler, E. Sabbi, A. Nota, M. Zamani (ESA/Webb).}
\label{fig:zeidler2024}
\end{figure*}

We obtained imaging data using the Near Infrared Camera (NIRCam) on April 24, 2023 and the Mid-Infrared Instrument (MIRI) on July 20, 2023, on-board JWST (Program: GO-2662, P.I.: P. Zeidler). The NIRCam observations utilized four medium and four wide-band filters: F115W, F140M, F150W, F210M, F277W, F335M, F356W, and F480M and for the MIRI observations, we used the filters F770W, F1000W, F1130W, F1500W, and F2550W.
NIRCam mapped a 3 $\times$ 5.8 arcminute$^{2}$ area while MIRI covered a smaller region of 1.2 $\times$ 1.9 arcminute$^{2}$. Both observations encompass the entire cluster and its surrounding H II regions. Figure~\ref{fig:zeidler2024} shows a color image of the NGC~602 using all NIRCam data and MIRI-F1000W. For further details on the NIRCam observations, refer to \cite{Zeidler2024}.

The MIRI observations were taken with a 2~$\times$~2 mosaic and a 4-POINT-SETS dither pattern. Exposure times for F770W, F1000W, and F1500W were 333.1 s per dither, with 30 groups/integration to achieve S/N~$\ge$~5. Due to the relatively narrower bandwidth of F1130W, it required 444.1~s with 40 groups, while F2550W had four integrations of 20 groups, totaling 888.1 s due to thermal noise limitation of the filter. Background observations were taken for each filter to ensure accurate sky subtraction. 

We reduced the MIRI observations using  the JWST data reduction pipeline (v1.12.5; \citealp{Bushouse2023}; CRDS context jwst\_1263.pmap), executing all standard pipeline steps (calwebb.detector1, calwebb.image2 and calwebb.image3). We aligned all observations to the NIRCam photometric catalog of \citet{Zeidler2024}. Detailed descriptions of the MIRI reduction and calibration process will be provided in the data paper (Meena 2025c, in preparation).

\section{Analysis and Results}\label{sec:analysis}

\subsection{Mid-IR Photometric Detection of Protocluster Candidate}\label{subsec:detect}
\begin{figure*}[ht!]
\centering
\includegraphics[width=\textwidth]{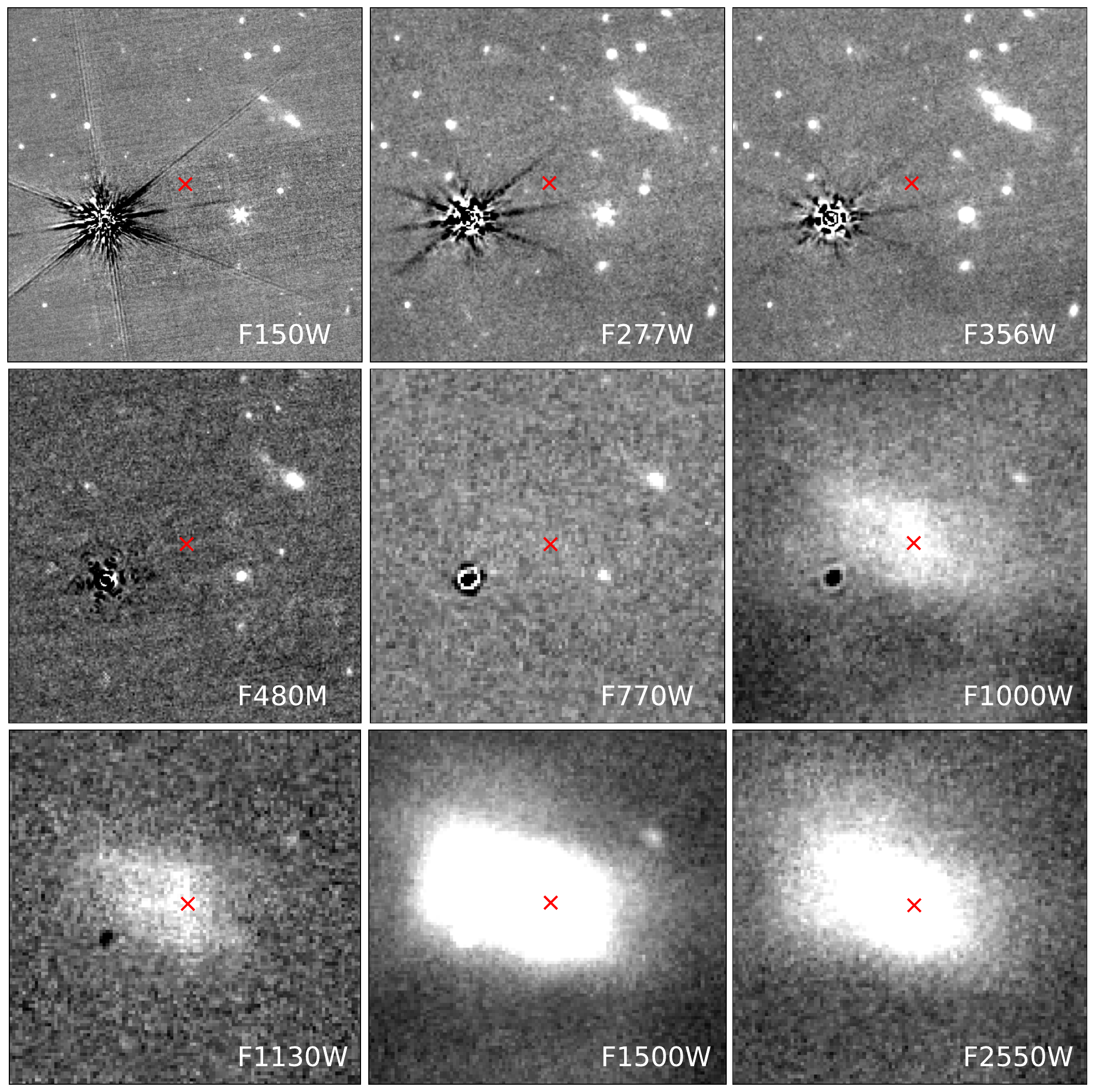}
\caption{Each panel presents a 12\arcsec\ $\times$ 12\arcsec\ cutout of the region surrounding the Sk183, spanning from wide-band NIRCam filters to MIRI filters. This progression highlights the gradual emergence of the extended emission as we move to longer wavelengths. In the NIRCam images, it remains undetected, while in the mid-IR, particularly in MIRI F1000W and longer-wavelength filters, the emission from MZS-1 becomes increasingly prominent. This trend suggests that this object is deeply embedded and heavily obscured at shorter wavelengths, with its emission peaking in the mid-IR. The apparent fainter nature of the F1130W filter is due to its relatively narrow bandwidth, which is less than half the width of other MIRI wide-band filters. This reduced bandwidth results in lower integrated flux compared to adjacent filters.} 
\label{fig:blob_nircam_miri}
\end{figure*}

While analyzing the MIRI F1000W image of NGC~602, we identified a region of enhanced mid-IR emission (R.A.~=~22.3495707\arcdeg, Dec.~=~-73.5541959\arcdeg, ICRS) in the northeastern part of the cluster center, located near the massive (46~$M_{\odot}$; \citealp{Evans2012}) O3-type star Sk183 \citep{Sanduleak1969}. We refer to this feature as MZS-1. Interestingly, MZS-1 has no detectable counterpart in any of the NIRCam filters and appears extremely faint in the shorter-wavelength MIRI F770W image.

A cutout of the region surrounding MZS-1 is shown in Figure~\ref{fig:blob_nircam_miri}, highlighting its emergence in the mid-IR. As seen in the image, emission from this region is completely absent at shorter wavelengths and only becomes apparent at wavelengths $\geq$10~$\micron$. The source appears elongated in all MIRI bands where it is detected, reaching an extent of up to approximately 8\arcsec~in the F1500W image (centered at 15~$\micron$), where the emission is strongest. Overall, the structure has an elliptical shape, with a major axis of $\sim$8\arcsec, a minor axis of $\sim$4\arcsec, and a position angle (PA) of $\sim$320$^\circ$ measured clockwise from North. Our initial observations suggest that the emission in this region is dominated by thermal radiation from warm dust or deeply embedded sources, which are heavily obscured at shorter wavelengths.

The massive early type O3 star Sk183 is extremely bright in both the optical and near-IR (V= 13.82$\pm$0.01 mag, \citealp{Massey2002}; J = 14.426$\pm$0.029 mag, \citealp{Evans2012}), which has a significant point spread function (PSF) in the NIRCam images. While NIRCam has a relatively narrower PSF compared to MIRI, the extreme brightness of Sk183 in the near-IR results in prominent diffraction spikes and extended PSF wings, potentially obscuring any faint nearby emission, such as that from MZS-1, if present. To ensure that MZS-1 is not hidden beneath the stellar PSF of Sk183, we performed PSF subtraction to remove the star's contribution.

We used WebbPSF \citep{Perrin2015} to generate model PSFs corresponding to the specific NIRCam filters and observational setups. The models were aligned to the exact coordinates of Sk183, scaled to match its observed fluxes (in each filter), and rotated to account for the telescope's V3 PA during the observation. We then subtracted these modeled PSFs from the observations. 

We also performed PSF subtraction in the MIRI F770W, F1000W, and F1130W images. In the F1500W and F2000W bands, however, the emission from MZS-1 dominates over Sk183, making PSF subtraction impractical in these data. The resulting PSF-subtracted NIRCam and MIRI images are shown in Figure~\ref{fig:blob_nircam_miri}. As seen in the figure, no residual emission is detected near Sk183 in any of the NIRCam filters. However, the extended emission from MZS-1 is clearly evident in the MIRI bands, particularly at longer wavelengths. We also created stacked NIRCam images by combining the four short-wavelength channels and, separately, the four long-wavelength channels to improve sensitivity. Neither of the resulting master images revealed any detectable counterpart at the location of MZS-1. 

The absence of MZS-1 in the near-IR and its prominent detection only at mid-IR wavelengths suggest it harbors a deeply embedded structure, likely a very early-stage YSO. The strong, spatially extended emission observed in the MIRI bands suggests that the emission does not come from a single source rather a group of embedded sources—potentially a protostar-cluster in formation.

\subsection{Two-dimensional Flux Map}\label{subsec:multi}

As discussed above, the elongated morphology of MZS-1 suggests that the mid-IR emission likely arises from multiple embedded YSOs, consistent with a collection of protostellar cores. This interpretation is further supported by the two-dimensional flux map in the MIRI F1130W band. Due to its relatively narrow bandwidth, F1130W provides enhanced contrast, allowing for a more detailed examination of the emission morphology. 

\begin{figure}[ht!]
\centering
\includegraphics[width=0.45\textwidth]{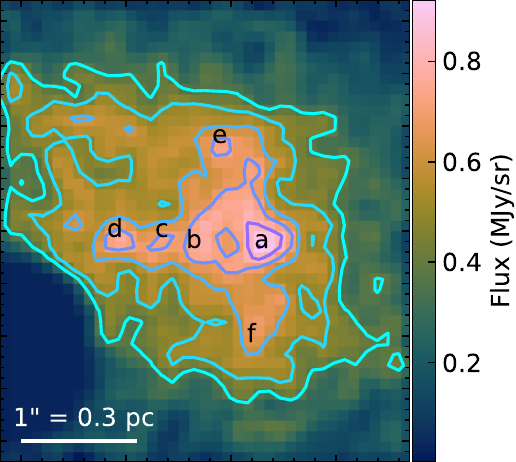}
\caption{F1130W image of MZS-1. The fainter emission in this band compared to the other MIRI filters allows for a clearer view of the morphology. Flux contours are overlaid at levels exceeding 10$\sigma$ above the background flux, following a Gaussian scaling over the image smoothed using a 2×2 box kernel. The several flux peaks 
(named MZS-1-`a', `b', `c', `d', `e', `f') are potentially associated with six distinct protostellar sources in MZS-1.} 
\label{fig:multipeaks}
\end{figure}

We generated a set of flux contours around MZS-1 using F1130W image as shown in Figure~\ref{fig:multipeaks}, which reveals at least six bright peaks corresponding to six potential protostellar sources.

For the contour analysis, we first estimated the background noise level by computing the standard deviation ($\sigma_{\mathrm{bg}}$) of the flux in a relatively empty, source-free region of the field 
(R.A. = 22.3849107\arcdeg, DEC. = -73.5476766\arcdeg, ICRS). 
Using this value as a reference, we then generated a set of flux contours. The outermost contour shown in Figure~\ref{fig:multipeaks} corresponds to a flux level of 10$\sigma_{\mathrm{bg}}$, ensuring that all detected features are well above the background noise.

Additionally, to reduce stellar contamination in our contour analysis, we masked out Sk183 and two nearby point sources located close to the emission structure. Circular masks were manually placed over each of these sources, with radii selected to fully enclose their PSF wings and minimize any residual flux that could affect the analysis of MZS-1.

\subsection{Spectral Energy Distribution}\label{subsec:sed_fit}

\begin{figure}[ht!]
\includegraphics[width=0.475\textwidth]{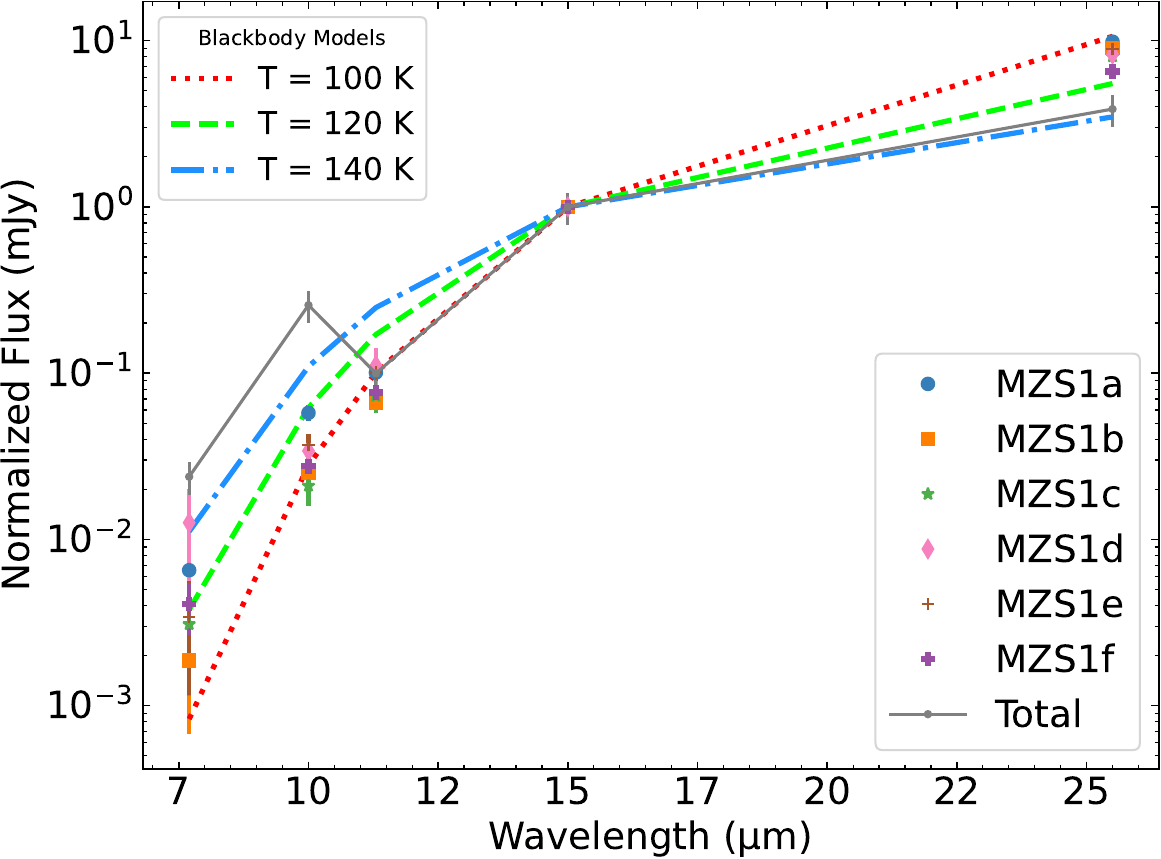}
\caption{SEDs generated for each core identified in Figure~\ref{fig:multipeaks}, derived using PSF photometry. Measured fluxes for each core are shown using distinct marker styles and colors, as indicated in the legend (bottom right). The gray line shows the total flux measured in a 8\arcsec $\times$ 4\arcsec~elliptical aperture around MZS-1. All fluxes are normalized to flux in F1500W. While individual cores vary in brightness, their overall SED shapes are comparable. Overlaid are blackbody curves for three temperatures: 100 K (red), 120 K (green), and 140 K (blue). The error bars correspond to flux uncertainties derived from the PSF-fitting procedure.}
\label{fig:sed_contours}
\end{figure}

We performed PSF photometry on the six compact flux peaks identified in Figure~\ref{fig:multipeaks}, as well as aperture photometry using Photutils \citep{larry_bradley_2024_13989456} on the integrated emission of MZS-1 of an 8\arcsec\ $\times$ 4\arcsec\ elliptical region. 

We estimated the initial centroids (x and y positions) of the peaks (potential protostellar cores) through visual identification in Figure~\ref{fig:multipeaks}. For the PSF photometry in F1130W, which provides the highest contrast, we allowed the centroids to vary within $\pm$3 pixels of the initial guess. To test the robustness of this procedure, we selected the brightest and faintest peaks in F1130W, perturbed their initial positions randomly within a 1$\times$1 pixel box around the visually identified location, and repeated the PSF photometry allowing centroid variation from those initial guesses. In all cases, the fitted centroids converged to consistent positions, with a standard deviation of $\sim$0.15 pixels for the brightest source and $\sim$0.5 pixels for the faintest source. The corresponding variation in flux was about 7$\%$ for the brightest source and 14$\%$ for the faintest source.
The fitted centroids obtained from F1130W were then fixed for the PSF photometry in the other filters. The resulting SEDs, constructed from the PSF-fitted fluxes of the individual cores together with the integrated flux of MZS-1, are shown in Figure~\ref{fig:sed_contours}.

While the total flux levels vary slightly among the cores, the overall shape of the SEDs is broadly consistent among all them. Each SED rises steeply toward longer wavelengths and peaks at 25.5~$\micron$. It is important to note that these identified protostellar cores maybe unresolved from other, potentially fainter sources.

To characterize the thermal properties of the emitting sources, we overlaid single-temperature blackbody curves on the SEDs. These illustrative models suggest characteristic temperatures in the range of 100-140 K. We emphasize that these blackbody curves are not formal fits, but rather serve to provide a first-order estimate of the blackbody temperatures for the observed mid-IR SEDs.

Based on their inferred cold temperatures, rising SEDs, and non-detections at shorter wavelengths, the identified cores within MZS-1 indicate potential presence of early-stage protostars—potentially Class 0/I YSOs—embedded in a nascent protocluster.

\subsection{YSO Modeling}\label{subsec:yso_fit}

To further characterize the properties of MZS-1 and its protostellar cores, we fit the observed SEDs using the T. P. Robitaille (2017) models for YSOs, convolved with JWST filters \citep{Robitaille2017, Richardson2024}. These synthetic YSO models cover a broad range of YSO evolutionary stages, from deeply embedded protostars to pre-main-sequence stars. The models consider a stellar core with stellar properties, including mass, luminosity, and temperature, as well as the accretion disk and surrounding envelope of gas and dust. For details of each model, see Table~2 in \cite{Robitaille2017}.

Since no emission is detected in any of the NIRCam filters and the region is contaminated by diffraction spikes from Sk183, we estimated an upper limit for the near-IR flux based on the local background level. Specifically, we measured the mean background flux within a 0.5\arcsec\ aperture centered on a nearby, source-free region of the field (R.A. = 22.3849107\arcdeg, DEC. = -73.5476766\arcdeg, ICRS).

To estimate the physical properties of the candidate YSOs within MZS-1, we utilized the full grid of model sets from Table~2 of \citet{Robitaille2017}. We allowed for a broad range of extinction values ($A_V = 0$–100) and constrained the distance to 61–63 kpc, reflecting the location of MZS-1 within the Magellanic Stream, which has an estimated thickness of about 2~kpc \citep{Muller2003}. Additionally, since no optical extinction (besides the average dust extinction of the cluster) is observed toward Sk183, MZS-1 is assumed to lie at or beyond the position of Sk183 along the line of sight.

As noted by \citet{Robitaille2017}, the goal of YSO fitting is not to produce fully realistic models, but rather to explore a broad parameter space using simplified components. To determine the most likely physical parameters of our sources, we adopt the Bayesian model selection approach described in their work, rather than relying solely on the single best-fit $\chi^2$ value.

In this method, each model set (i.e., a grid of models with the same physical components) is evaluated based on the number of models that provide good fits to the data, defined as those with $\chi^2 - \chi^2_\mathrm{best} < 9$. The relative likelihood of each model set is then given by the ratio $N_\mathrm{good} / N$, where $N$ is the total number of models in the set.

Finally, because not all parameter combinations in these models are physically plausible \citep{Robitaille2017}, we restricted the final selected models so that the stellar radii and masses (derived from stellar luminosities) yield surface gravities in the range 3.5 $\le~log~g~\le$ 5 , which are typical for YSOs/PMS stars \citep{Kounkel2018, Lopez-Valdivia2021}. We applied this approach to our SED fits of each protostellar core, evaluating all model sets described in Table 2 of~\citet{Robitaille2017}. For the six cores, the best fits were obtained with either the `spubhmi' model set (central star + disk + Ulrich-type infalling envelope + bipolar cavity + ambient ISM) or the `spu-hmi model set (central star + disk + Ulrich-type envelope + ambient medium), provided they yielded the highest likelihood while also satisfying the $\chi^2$ and $log~g$ criteria.

\begin{figure}[ht!]
\centering
\includegraphics[width=0.45\textwidth]{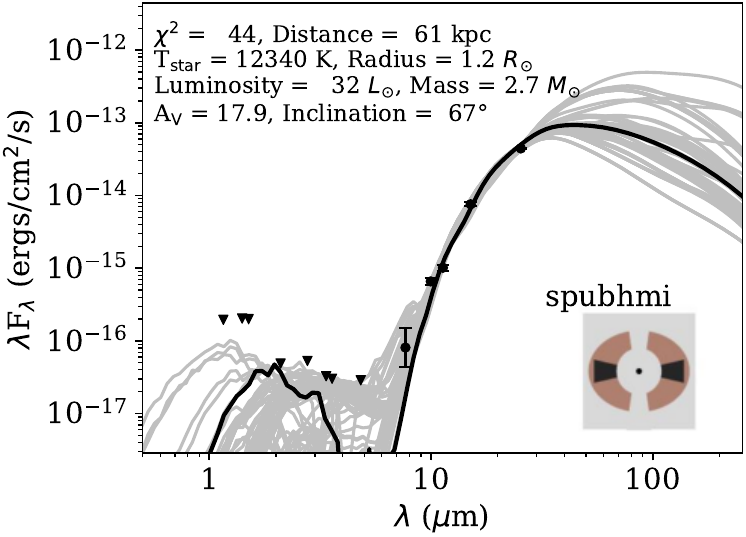} 
\caption{YSO fits for MZS-1a using the \cite{Robitaille2017} model grid `spubhmi'. A cartoon representation of the YSO model configuration is shown in the insets. The observed MIRI fluxes are plotted as black circles, while the upper limits for the NIRCam filters (obtained using the background region) are indicated by downward triangles. The error-bars on MIRI data-points represent the uncertainties in flux calculated from PSF fitting. The best-fit model is shown as the solid black line, and the corresponding output parameters are listed at the top left. Here the mass of the central star is derived using L = M$^{3.5}$ $L_{\odot}$. The SED fitting for rest of the candidate protostellar cores are given in the Appendix (\S~\ref{app:ysomodel}).}
\label{fig:sed_fits}
\end{figure}

An example SED fit for a single YSO (MZS1-a) is shown in Figure~\ref{fig:sed_fits}, with the associated best-fit parameters listed in the legend. SED fits for the remaining protostellar candidates are presented in Appendix~\ref{fig:app1_yso}. Table~\ref{tab1:YSO_params} summarizes the derived physical parameters for all cores, reporting the best-fit values along with lower and upper limits based on the distribution of all good-fit models within the selected model set.

Across all protostellar cores, the inferred stellar masses are below $\sim$3~$M_{\odot}$, and stellar radii are within $\sim$1.2~$R_{\odot}$, while extinctions ($A_V$) ranges from 17-41 $mag$. The presence of a dense Ulrich envelope, high extinction, low blackbody temperatures, and the overall SED shapes suggest that these objects are likely low-mass, Stage I (Class 0/I) protostars.

However, it is important to note that these YSO models do not include PAH emission, which may be present and thus affect the 7.7~$\micron$ and 11.3~$\micron$ MIRI data. Consequently, the derived model parameters—and by extension, the assigned YSO stages—should be interpreted with caution. Nonetheless, these fits confirm that the sources are low-mass and likely fell below Spitzer’s detection threshold. The IRAC 8 $\micron$ sensitivity ($\sim$0.5 mJy) corresponds to a detection limit of approximately 2–4~$M_\odot$ for embedded Class I YSOs \citep{Carlson2011}, which explains their absence in earlier Spitzer studies.

\begin{deluxetable*}{lcccchcchhhh}[ht!]
\setlength{\tabcolsep}{0.15in}
\renewcommand{\arraystretch}{1.4}
\caption{YSO model fit parameters}\label{tab1:YSO_params}
\tablehead{
  \colhead{Source} &
  \colhead{Model} &
  \colhead{$R_{\star}$ ($R_{\odot}$)} &
  \colhead{$T_{\star}$ (K)} &
  \colhead{$L_{\star}$ ($L_{\odot}$)} &
  \nocolhead{Inclination ($\arcdeg$)} &
  \colhead{$M_{\star}$ ($M_{\odot}$)} &
  \colhead{log $g$} &
  \nocolhead{A$_V$ (mag)} & 
  \nocolhead{info\_av} &  
  \nocolhead{info\_chi2} &
  \nocolhead{info\_sc}
}
\startdata
MZS1a & spubhmi & $1.2^{+2.8}_{-0}$ & $12300^{+860}_{-0}$ & $31.7^{+429}_{-0}$ & $67.7^{+19.7}_{-0}$ & $2.7^{+3.1}_{-0}$ & $4.7^{+0}_{-0.7}$ & $1.9e+03^{+0}_{-1.47e+03}$ & $17.9^{+17.2}_{-0}$ & $44.4^{+2.65}_{-0}$ & $1.79^{+0}_{-0}$ \\
MZS1b & spu-hmi & $1.0^{+0.8}_{-0}$ & $12500^{+4290}_{-1530}$ & $21.6^{+105}_{-0}$ & $88.8^{+0}_{-5.19}$ & $2.4^{+1.6}_{-0}$ & $4.8^{+0}_{-0.4}$ & $22.4^{+9.52e+03}_{-0}$ & $43.5^{+0}_{-13.5}$ & $21.7^{+21.5}_{-0}$ & $1.79^{+0}_{-0.00284}$ \\
MZS1c & spubhmi & $0.8^{+3.8}_{-0}$ & $13000^{+5120}_{-6690}$ & $17.3^{+318}_{-6.1}$ & $85.3^{+4.69}_{-83.4}$ & $2.3^{+3.0}_{-0.3}$ & $4.9^{+0.0}_{-1.3}$ & $4.76^{+2.86e+06}_{-3.88}$ & $40.5^{+43.6}_{-40.5}$ & $5.3^{+28.9}_{-0}$ & $1.79^{+0}_{-0.00284}$ \\
MZS1d & spu-hmi & $1.0^{+2.6}_{-0}$ & $14600^{+0}_{-2990}$ & $42.3^{+274}_{-13.2}$ & $85.6^{+0.895}_{-4.48}$ & $2.9^{+2.3}_{-0.3}$ & $4.8^{+0}_{-0.8}$ & $1.64e+03^{+0}_{-1.63e+03}$ & $30.1^{+7.93}_{-1.96}$ & $10.4^{+5.91}_{-0}$ & $1.79^{+0}_{-0.00284}$ \\
MZS1e & spubhmi & $1.2^{+2.1}_{-0.4}$ & $9500^{+9780}_{-894}$ & $12^{+324}_{-0.74}$ & $62.1^{+27.9}_{-36.9}$ & $2.0^{+3.2}_{-0.1}$ & $4.5^{+0.4}_{-0.7}$ & $21.6^{+2.86e+06}_{-16.9}$ & $33.3^{+20.5}_{-33.3}$ & $11.2^{+21.9}_{-0}$ & $1.79^{+0.00284}_{-0}$ \\
MZS1f & spubhmi & $0.8^{+2.5}_{-0}$ & $13000^{+610}_{-4650}$ & $17.3^{+318}_{-6.1}$ & $85.3^{+0}_{-55.8}$ & $2.3^{+3.0}_{-0.3}$ & $4.97^{+0}_{-1.0}$ & $4.76^{+6.72e+04}_{-0}$ & $32.1^{+21.5}_{-28.6}$ & $18.9^{+25.9}_{-0}$ & $1.79^{+0}_{-0.00284}$ \\
\enddata
\tablecomments{Stellar mass M$_{\star}$ is estimated using $L \sim M^{3.5}$.}
\end{deluxetable*}

\section{Discussion and Conclusion}\label{sec:discussion}

We identify a candidate Class 0/I protocluster, which we designate as MZS-1, in JWST/MIRI imaging of NGC 602. It lies in projection near the bright O3 star Sk183 and spans approximately 2.4~pc at the distance of the SMC. However, the actual three-dimensional distance between Sk183 and MZS-1 is unknown. The structure contains at least six protostellar cores, each with a mass $\leq$ $3$~$M_{\odot}$.

Previously, using IR photometry of Sk183, using the Two Micron All Sky Survey (2MASS), Wide-field Infrared Survey Explorer (WISE), and Spitzer's InfraRed Array Camera (IRAC), \citet{Evans2012} had found a mid-IR excess in the longer-wavelength WISE data. By further inspecting this region with Spitzer/MIPS 24$\micron$ imaging, they found an extended morphology with the emission peak slightly offset from Sk183, postulating that the excess did not originate from Sk183 itself but rather from nearby dust emission or unresolved shocks. However, due to the limited resolution of WISE and Spitzer, this feature remained largely unconstrained. Notably, MZS-1 is also undetected in far-IR Herschel observations \citep{Seale2014}.

Our JWST observations now clearly resolve this structure—designated MZS-1—in unprecedented detail. MZS-1 appears as an extended mid-IR feature northeast of Sk183, most prominent in the MIRI F1500W image, and spans up to $\sim$8\arcsec~(2.4~pc at the distance of the SMC). Its complete absence in all NIRCam filters strongly suggests the presence of deeply embedded sources that are heavily obscured at shorter wavelengths.
Furthermore, the detection of multiple emission peaks within MZS-1 implies the presence of several protostellar sources. YSO modeling of the six brightest peaks identified in the 2D flux maps using the grid of \citet{Robitaille2017} indicates stellar masses below $\sim$3~$M_\odot$. This low mass range helps explain why these sources are absent from earlier Spitzer-based YSO catalogs (e.g., \citealt{Carlson2007, Carlson2011}), which had a detection limit of $\sim$2–4~$M_\odot$ at the distance of the Magellanic Clouds \citep{Meixner2006, Sewilo2013}.

The best-fit models for YSOs includes a thick dusty envelope, consistent with highly embedded Stage I YSOs. This is supported by the observed positive spectral index in the IR and cold blackbody-like temperatures of $\sim$100–140~K inferred from the SEDs, characterizing low mass young protostellar cores (likely stage~I) still embedded in their natal envelopes.

Given that we detect the most massive components of the protocluster (associated with six flux peaks), we can estimate the total stellar mass of the protocluster by applying an initial mass function (IMF) to the YSO-derived masses. Assuming a Salpeter IMF \citep{Salpeter1955} with a lower mass cutoff of 0.1~$M_\odot$, the six detected sources in the 2-3~$M_\odot$ range (Table~\ref{tab1:YSO_params}) represent approximately 5$\%$ of the total stellar mass. This implies a total protocluster mass of $\approx$300~$M_\odot$, consistent with MZS-1 being a low-mass protocluster forming in a low-density environment where no YSOs more massive than $\sim$3~$M_\odot$ are present or detected.

High-resolution ALMA observations by \citet{ONeill2022} identified 110 molecular clumps in NGC~602, but no significant CO(2–1) emission is detected near MZS-1. In low-metallicity environments like the SMC, CO emission is often faint, and the CO-to-H$_2$ conversion factor is not reliable \citep{Israel1986, Israel1997, Bolatto2013}. Although \citet{ONeill2022} applied corrections for “CO-dark” gas, CO remains undetected at MZS-1's location. This is likely due to photodissociation of CO by far-ultraviolet radiation, while H$_2$, being more self-shielding, survives \citep{Madden2006, Wolfire2010, Gordon2011, Madden2020}. Additionally, if MZS-1 is in an early evolutionary stage, CO excitation may be inefficient, or the emission may fall below ALMA’s sensitivity. Therefore, the absence of CO does not rule out cold, dense gas but highlights the need for alternative tracers or deeper submillimeter observations to probe MZS-1's nature.

There is evidence that far-UV radiation from massive stars can influence, and in some cases trigger, star formation in their surroundings \citep{HLee2005,Ellerbroek2013,Luisi2021}. In NGC 602, studies of the YSO and PMS populations suggest that the central massive stars may have triggered star formation along the inner ridge of the H II region \citep{Carlson2007, Carlson2011, Gouliermis2012}, though other works have argued against this interpretation \citep{DeMarchi2013, ONeill2022}. If MZS-1 is associated with NGC 602, its formation may have been affected by the ionizing output of the nearby O3 star Sk183, which dominates the ionizing flux in the region \cite{Evans2012, Ramachandran2019}. \citet{ONeill2022} proposed that the lack of CO emission near Sk183 could result from complete photodissociation of molecular gas by its ionizing photons. The existence of a cold, embedded source like MZS-1 argues against total disruption. A likely explanation is that MZS-1 lies at a much larger physical distance from Sk183 than implied by its projected separation. Nevertheless, without distance measurements it is difficult to assess whether and how the O3 star influences the protocluster, or has played any role in, its formation.

Before identifying MZS-1 as a group of stage I YSOs, we also considered alternative scenarios to explain the emission from this region. For example, given its size, morphology, and the absence of emission at shorter wavelengths, we rule out the possibility that it is a background galaxy. MZS-1 has a relatively large angular size ($\sim$8\arcsec), which would correspond to an unusually large physical size ($\ge$30 kpc) if located at moderate redshifts (z $\ge$ 0.25). While such extended galaxies are not impossible, they are rare \citep{vanderWel2014}, and MZS-1 would be significantly larger than other galaxies observed in the background of NGC 602. Additionally, its extremely red mid-IR color ($>$6 mag between 4$\micron$ and 12$\micron$) is inconsistent with known populations of ultra-luminous infrared galaxies or active galactic nuclei (AGN), which typically exhibit much shallower mid-IR slopes \citep{Dey2024}. Finally, the observed SED of MZS-1 resembles a cold blackbody, in contrast to the SEDs associated with galaxies and AGN \citep{Stern2005}.

The SMC and LMC are known to harbor isolated compact H II regions, including Low Excitation Blobs (LEBs) and High Excitation Blobs (HEBs) \cite{Heydari-Malayeri1990, Heydari-Malayeri2010}. HEBs, in particular, exhibit strong mid-IR emission \citep{Selier2011, Stephens2017, Charmandaris2008}. However, these blobs also show prominent optical high-ionization lines like H-alpha and H-beta \citep{Testor2014}, which MZS-1 lacks, ruling out the possibility of it being an LEB or HEB.

Finally, the absence of scattered light in optical wavelengths \citep{Draine2003,Gordon2004, Hensley2023} rules out the presence of cold ISM or large dust grains. 

Our findings emphasize the need for further data to fully understand MZS-1. Future mid-IR spectroscopy and deep submillimeter observations will be crucial for constraining its physical and chemical properties and the surrounding environment, shedding light on the early epoch of low-mass star-cluster formation in metal-poor regime.


\begin{acknowledgments}
This work is based on observations made with the NASA/ESA/CSA James Webb Space Telescope. The data were obtained from the Mikulski Archive for Space Telescopes at the Space Telescope Science Institute, which is operated by the Association of Universities for Research in Astronomy, Inc., under NASA contract NAS 5-03127 for JWST. These observations are associated with program $\#$2662. The JWST data presented in this article were obtained from the Mikulski Archive for Space Telescopes (MAST) at the Space Telescope Science Institute. The specific observations used in this work can be access via \href{https://archive.stsci.edu/doi/resolve/resolve.html?doi=10.17909/x2h4-kv56}{doi:10.17909/x2h4-kv56} for NIRCam data and \href{https://archive.stsci.edu/doi/resolve/resolve.html?doi=10.17909/54tf-jq46}{doi:10.17909/54tf-jq46} for MIRI data. 

E.S. is supported by the international Gemini Observatory, a program of NSF NOIRLab, which is managed by the Association of Universities for Research in Astronomy (AURA) under a cooperative agreement with the U.S. National Science Foundation, on behalf of the Gemini partnership of Argentina, Brazil, Canada, Chile, the Republic of Korea, and the United States of America. OCJ has received funding from an STFC Webb fellowship.
This research made use of Photutils, an Astropy package for
detection and photometry of astronomical sources \citep{larry_bradley_2024_13989456}.

\end{acknowledgments}
\facilities{\textit{JWST} (NIRCam, MIRI)}

\software{SAOImage DS9 \citep{ds92000}, Python (\citealp{VanRossum2009}, \url{https://www.python.org}), Astropy \citep{astropy:2013, astropy:2018, astropy:2022}}

\bibliography{refs}{}

\begin{thebibliography}{}
\expandafter\ifx\csname natexlab\endcsname\relax\def\natexlab#1{#1}\fi
\providecommand{\url}[1]{\href{#1}{#1}}
\providecommand{\dodoi}[1]{doi:~\href{http://doi.org/#1}{\nolinkurl{#1}}}
\providecommand{\doeprint}[1]{\href{http://ascl.net/#1}{\nolinkurl{http://ascl.net/#1}}}
\providecommand{\doarXiv}[1]{\href{https://arxiv.org/abs/#1}{\nolinkurl{https://arxiv.org/abs/#1}}}

\bibitem[{{Andre} {et~al.}(2000){Andre}, {Ward-Thompson}, \& {Barsony}}]{Andre2000}
{Andre}, P., {Ward-Thompson}, D., \& {Barsony}, M. 2000, in Protostars and Planets IV, ed. V.~{Mannings}, A.~P. {Boss}, \& S.~S. {Russell}, 59, \dodoi{10.48550/arXiv.astro-ph/9903284}

\bibitem[{{Astropy Collaboration} {et~al.}(2013){Astropy Collaboration}, {Robitaille}, {Tollerud}, {Greenfield}, {Droettboom}, {Bray}, {Aldcroft}, {Davis}, {Ginsburg}, {Price-Whelan}, {Kerzendorf}, {Conley}, {Crighton}, {Barbary}, {Muna}, {Ferguson}, {Grollier}, {Parikh}, {Nair}, {Unther}, {Deil}, {Woillez}, {Conseil}, {Kramer}, {Turner}, {Singer}, {Fox}, {Weaver}, {Zabalza}, {Edwards}, {Azalee Bostroem}, {Burke}, {Casey}, {Crawford}, {Dencheva}, {Ely}, {Jenness}, {Labrie}, {Lim}, {Pierfederici}, {Pontzen}, {Ptak}, {Refsdal}, {Servillat}, \& {Streicher}}]{astropy:2013}
{Astropy Collaboration}, {Robitaille}, T.~P., {Tollerud}, E.~J., {et~al.} 2013, \aap, 558, A33, \dodoi{10.1051/0004-6361/201322068}

\bibitem[{{Astropy Collaboration} {et~al.}(2022){Astropy Collaboration}, {Price-Whelan}, {Lim}, {Earl}, {Starkman}, {Bradley}, {Shupe}, {Patil}, {Corrales}, {Brasseur}, {N{"o}the}, {Donath}, {Tollerud}, {Morris}, {Ginsburg}, {Vaher}, {Weaver}, {Tocknell}, {Jamieson}, {van Kerkwijk}, {Robitaille}, {Merry}, {Bachetti}, {G{"u}nther}, {Aldcroft}, {Alvarado-Montes}, {Archibald}, {B{'o}di}, {Bapat}, {Barentsen}, {Baz{'a}n}, {Biswas}, {Boquien}, {Burke}, {Cara}, {Cara}, {Conroy}, {Conseil}, {Craig}, {Cross}, {Cruz}, {D'Eugenio}, {Dencheva}, {Devillepoix}, {Dietrich}, {Eigenbrot}, {Erben}, {Ferreira}, {Foreman-Mackey}, {Fox}, {Freij}, {Garg}, {Geda}, {Glattly}, {Gondhalekar}, {Gordon}, {Grant}, {Greenfield}, {Groener}, {Guest}, {Gurovich}, {Handberg}, {Hart}, {Hatfield-Dodds}, {Homeier}, {Hosseinzadeh}, {Jenness}, {Jones}, {Joseph}, {Kalmbach}, {Karamehmetoglu}, {Ka{l}uszy{'n}ski}, {Kelley}, {Kern}, {Kerzendorf}, {Koch}, {Kulumani}, {Lee}, {Ly}, {Ma}, {MacBride}, {Maljaars}, {Muna}, {Murphy}, {Norman}, {O'Steen},
  {Oman}, {Pacifici}, {Pascual}, {Pascual-Granado}, {Patil}, {Perren}, {Pickering}, {Rastogi}, {Roulston}, {Ryan}, {Rykoff}, {Sabater}, {Sakurikar}, {Salgado}, {Sanghi}, {Saunders}, {Savchenko}, {Schwardt}, {Seifert-Eckert}, {Shih}, {Jain}, {Shukla}, {Sick}, {Simpson}, {Singanamalla}, {Singer}, {Singhal}, {Sinha}, {Sip{H{o}}cz}, {Spitler}, {Stansby}, {Streicher}, {{{S}}umak}, {Swinbank}, {Taranu}, {Tewary}, {Tremblay}, {Val-Borro}, {Van Kooten}, {Vasovi{'c}}, {Verma}, {de Miranda Cardoso}, {Williams}, {Wilson}, {Winkel}, {Wood-Vasey}, {Xue}, {Yoachim}, {Zhang}, {Zonca}, \& {Astropy Project Contributors}}]{astropy:2022}
{Astropy Collaboration}, {Price-Whelan}, A.~M., {Lim}, P.~L., {et~al.} 2022, \apj, 935, 167, \dodoi{10.3847/1538-4357/ac7c74}

\bibitem[{{Bolatto} {et~al.}(2013){Bolatto}, {Wolfire}, \& {Leroy}}]{Bolatto2013}
{Bolatto}, A.~D., {Wolfire}, M., \& {Leroy}, A.~K. 2013, \araa, 51, 207, \dodoi{10.1146/annurev-astro-082812-140944}

\bibitem[{Bradley {et~al.}(2024)Bradley, Sip{\H o}cz, Robitaille, Tollerud, Vin{\'{\i}}cius, Deil, Barbary, Wilson, Busko, Donath, G{\"u}nther, Cara, Lim, Me{\ss}linger, Conseil, Burnett, Bostroem, Droettboom, Bray, Bratholm, Ginsburg, Jamieson, Barentsen, Craig, Morris, Perrin, Rathi, Pascual, \& Georgiev}]{larry_bradley_2024_13989456}
Bradley, L., Sip{\H o}cz, B., Robitaille, T., {et~al.} 2024, astropy/photutils: 2.0.2, 2.0.2,  Zenodo, \dodoi{10.5281/zenodo.13989456}

\bibitem[{{Bushouse} {et~al.}(2023){Bushouse}, {Eisenhamer}, {Dencheva}, {Davies}, {Greenfield}, {Morrison}, {Hodge}, {Simon}, {Grumm}, {Droettboom}, {Slavich}, {Sosey}, {Pauly}, {Miller}, {Jedrzejewski}, {Hack}, {Davis}, {Crawford}, {Law}, {Gordon}, {Regan}, {Cara}, {MacDonald}, {Bradley}, {Shanahan}, {Jamieson}, {Teodoro}, {Williams}, \& {Pena-Guerrero}}]{Bushouse2023}
{Bushouse}, H., {Eisenhamer}, J., {Dencheva}, N., {et~al.} 2023, {JWST Calibration Pipeline}, 1.12.5,  Zenodo, \dodoi{10.5281/zenodo.10022973}

\bibitem[{Carlson {et~al.}(2007)Carlson, Sabbi, Sirianni, Hora, Nota, Meixner, Gallagher~III, Oey, Pasquali, Smith, Tosi, \& Walterbos}]{Carlson2007}
Carlson, L.~R., Sabbi, E., Sirianni, M., {et~al.} 2007, The Astrophysical Journal, 665, L109, \dodoi{10.1086/521023}

\bibitem[{{Carlson} {et~al.}(2011){Carlson}, {Sewi{\l}o}, {Meixner}, {Romita}, {Whitney}, {Hora}, {Cignoni}, {Sabbi}, {Nota}, {Sirianni}, {Smith}, {Gordon}, {Babler}, {Bracker}, {Gallagher}, {Meade}, {Misselt}, {Pasquali}, \& {Shiao}}]{Carlson2011}
{Carlson}, L.~R., {Sewi{\l}o}, M., {Meixner}, M., {et~al.} 2011, \apj, 730, 78, \dodoi{10.1088/0004-637X/730/2/78}

\bibitem[{{Charmandaris} {et~al.}(2008){Charmandaris}, {Heydari-Malayeri}, \& {Chatzopoulos}}]{Charmandaris2008}
{Charmandaris}, V., {Heydari-Malayeri}, M., \& {Chatzopoulos}, E. 2008, \aap, 487, 567, \dodoi{10.1051/0004-6361:200809662}

\bibitem[{Cignoni {et~al.}(2009)Cignoni, Sabbi, Nota, Tosi, Degl'Innocenti, Moroni, Angeretti, Carlson, Gallagher, Meixner, Sirianni, \& Smith}]{Cignoni2009}
Cignoni, M., Sabbi, E., Nota, A., {et~al.} 2009, Astronomical Journal, 137, 3668, \dodoi{10.1088/0004-6256/137/3/3668}

\bibitem[{{De Marchi} {et~al.}(2013){De Marchi}, {Beccari}, \& {Panagia}}]{DeMarchi2013}
{De Marchi}, G., {Beccari}, G., \& {Panagia}, N. 2013, \apj, 775, 68, \dodoi{10.1088/0004-637X/775/1/68}

\bibitem[{{Dey} {et~al.}(2024){Dey}, {Goyal}, {Ma{\l}ek}, \& {D{\'\i}az-Santos}}]{Dey2024}
{Dey}, S., {Goyal}, A., {Ma{\l}ek}, K., \& {D{\'\i}az-Santos}, T. 2024, \apj, 966, 61, \dodoi{10.3847/1538-4357/ad2c93}

\bibitem[{{Draine}(2003)}]{Draine2003}
{Draine}, B.~T. 2003, \araa, 41, 241, \dodoi{10.1146/annurev.astro.41.011802.094840}

\bibitem[{{Ellerbroek} {et~al.}(2013){Ellerbroek}, {Bik}, {Kaper}, {Maaskant}, {Paalvast}, {Tramper}, {Sana}, {Waters}, \& {Balog}}]{Ellerbroek2013}
{Ellerbroek}, L.~E., {Bik}, A., {Kaper}, L., {et~al.} 2013, \aap, 558, A102, \dodoi{10.1051/0004-6361/201321752}

\bibitem[{{Evans} {et~al.}(2012){Evans}, {Hainich}, {Oskinova}, {Gallagher}, {Chu}, {Gruendl}, {Hamann}, {H{\'e}nault-Brunet}, \& {Todt}}]{Evans2012}
{Evans}, C.~J., {Hainich}, R., {Oskinova}, L.~M., {et~al.} 2012, \apj, 753, 173, \dodoi{10.1088/0004-637X/753/2/173}

\bibitem[{{Fukui} {et~al.}(2020){Fukui}, {Ohno}, {Tsuge}, {Sano}, \& {Tachihara}}]{Fukui2020}
{Fukui}, Y., {Ohno}, T., {Tsuge}, K., {Sano}, H., \& {Tachihara}, K. 2020, arXiv e-prints, arXiv:2005.13750, \dodoi{10.48550/arXiv.2005.13750}

\bibitem[{{Garcia} {et~al.}(2021){Garcia}, {Evans}, {Bestenlehner}, {Bouret}, {Castro}, {Cervi{\~n}o}, {Fullerton}, {Gieles}, {Herrero}, {de Koter}, {Lennon}, {van Loon}, {Martins}, {de Mink}, {Najarro}, {Negueruela}, {Sana}, {Sim{\'o}n-D{\'\i}az}, {Sz{\'e}csi}, {Tramper}, {Vink}, \& {Wofford}}]{Garcia2021}
{Garcia}, M., {Evans}, C.~J., {Bestenlehner}, J.~M., {et~al.} 2021, Experimental Astronomy, 51, 887, \dodoi{10.1007/s10686-021-09785-x}

\bibitem[{{Gordon}(2004)}]{Gordon2004}
{Gordon}, K.~D. 2004, in Astronomical Society of the Pacific Conference Series, Vol. 309, Astrophysics of Dust, ed. A.~N. {Witt}, G.~C. {Clayton}, \& B.~T. {Draine}, 77, \dodoi{10.48550/arXiv.astro-ph/0309709}

\bibitem[{{Gordon} {et~al.}(2011){Gordon}, {Meixner}, {Meade}, {Whitney}, {Engelbracht}, {Bot}, {Boyer}, {Lawton}, {Sewi{\l}o}, {Babler}, {Bernard}, {Bracker}, {Block}, {Blum}, {Bolatto}, {Bonanos}, {Harris}, {Hora}, {Indebetouw}, {Misselt}, {Reach}, {Shiao}, {Tielens}, {Carlson}, {Churchwell}, {Clayton}, {Chen}, {Cohen}, {Fukui}, {Gorjian}, {Hony}, {Israel}, {Kawamura}, {Kemper}, {Leroy}, {Li}, {Madden}, {Marble}, {McDonald}, {Mizuno}, {Mizuno}, {Muller}, {Oliveira}, {Olsen}, {Onishi}, {Paladini}, {Paradis}, {Points}, {Robitaille}, {Rubin}, {Sandstrom}, {Sato}, {Shibai}, {Simon}, {Smith}, {Srinivasan}, {Vijh}, {Van Dyk}, {van Loon}, \& {Zaritsky}}]{Gordon2011}
{Gordon}, K.~D., {Meixner}, M., {Meade}, M.~R., {et~al.} 2011, \aj, 142, 102, \dodoi{10.1088/0004-6256/142/4/102}

\bibitem[{{Gouliermis} {et~al.}(2012){Gouliermis}, {Schmeja}, {Dolphin}, {Gennaro}, {Tognelli}, \& {Prada Moroni}}]{Gouliermis2012}
{Gouliermis}, D.~A., {Schmeja}, S., {Dolphin}, A.~E., {et~al.} 2012, \apj, 748, 64, \dodoi{10.1088/0004-637X/748/1/64}

\bibitem[{{Hensley} \& {Draine}(2023)}]{Hensley2023}
{Hensley}, B.~S., \& {Draine}, B.~T. 2023, \apj, 948, 55, \dodoi{10.3847/1538-4357/acc4c2}

\bibitem[{{Heydari-Malayeri} \& {Selier}(2010)}]{Heydari-Malayeri2010}
{Heydari-Malayeri}, M., \& {Selier}, R. 2010, \aap, 517, A39, \dodoi{10.1051/0004-6361/201014230}

\bibitem[{{Heydari-Malayeri} {et~al.}(1990){Heydari-Malayeri}, {van Drom}, \& {Leisy}}]{Heydari-Malayeri1990}
{Heydari-Malayeri}, M., {van Drom}, E., \& {Leisy}, P. 1990, \aap, 240, 481

\bibitem[{Hilditch {et~al.}(2005)Hilditch, Howarth, \& Harries}]{Hilditch2005}
Hilditch, R.~W., Howarth, I.~D., \& Harries, T.~J. 2005, Monthly Notices of the Royal Astronomical Society, 357, 304, \dodoi{10.1111/j.1365-2966.2005.08653.x}

\bibitem[{{Israel}(1997)}]{Israel1997}
{Israel}, F.~P. 1997, \aap, 328, 471, \dodoi{10.48550/arXiv.astro-ph/9709194}

\bibitem[{{Israel} {et~al.}(1986){Israel}, {de Graauw}, {van de Stadt}, \& {de Vries}}]{Israel1986}
{Israel}, F.~P., {de Graauw}, T., {van de Stadt}, H., \& {de Vries}, C.~P. 1986, \apj, 303, 186, \dodoi{10.1086/164065}

\bibitem[{{Kewley} \& {Kobulnicky}(2005)}]{Kewley2005}
{Kewley}, L., \& {Kobulnicky}, H.~A. 2005, in Astrophysics and Space Science Library, Vol. 329, Starbursts: From 30 Doradus to Lyman Break Galaxies, ed. R.~{de Grijs} \& R.~M. {Gonz{\'a}lez Delgado}, 307, \dodoi{10.1007/1-4020-3539-X_55}

\bibitem[{{Kounkel} {et~al.}(2018){Kounkel}, {Covey}, {Su{\'a}rez}, {Rom{\'a}n-Z{\'u}{\~n}iga}, {Hernandez}, {Stassun}, {Jaehnig}, {Feigelson}, {Pe{\~n}a Ram{\'\i}rez}, {Roman-Lopes}, {Da Rio}, {Stringfellow}, {Kim}, {Borissova}, {Fern{\'a}ndez-Trincado}, {Burgasser}, {Garc{\'\i}a-Hern{\'a}ndez}, {Zamora}, {Pan}, \& {Nitschelm}}]{Kounkel2018}
{Kounkel}, M., {Covey}, K., {Su{\'a}rez}, G., {et~al.} 2018, \aj, 156, 84, \dodoi{10.3847/1538-3881/aad1f1}

\bibitem[{{Lada}(1987)}]{Lada1987}
{Lada}, C.~J. 1987, in IAU Symposium, Vol. 115, Star Forming Regions, ed. M.~{Peimbert} \& J.~{Jugaku}, 1

\bibitem[{{Lee} {et~al.}(2005){Lee}, {Chen}, {Zhang}, \& {Hu}}]{HLee2005}
{Lee}, H.-T., {Chen}, W.~P., {Zhang}, Z.-W., \& {Hu}, J.-Y. 2005, \apj, 624, 808, \dodoi{10.1086/429122}

\bibitem[{Lee {et~al.}(2005)Lee, Rolleston, Dufton, \& Ryans}]{Lee2005}
Lee, J.~K., Rolleston, W.~R., Dufton, P.~L., \& Ryans, R.~S. 2005, Astronomy and Astrophysics, 429, 1025, \dodoi{10.1051/0004-6361:20041345}

\bibitem[{{L{\'o}pez-Valdivia} {et~al.}(2021){L{\'o}pez-Valdivia}, {Sokal}, {Mace}, {Kidder}, {Hussaini}, {Nofi}, {Prato}, {Johns-Krull}, {Oh}, {Lee}, {Park}, {Oh}, {Kraus}, {Kaplan}, {Llama}, {Mann}, {Kim}, {Gully-Santiago}, {Lee}, {Pak}, {Hwang}, \& {Jaffe}}]{Lopez-Valdivia2021}
{L{\'o}pez-Valdivia}, R., {Sokal}, K.~R., {Mace}, G.~N., {et~al.} 2021, \apj, 921, 53, \dodoi{10.3847/1538-4357/ac1a7b}

\bibitem[{{Luisi} {et~al.}(2021){Luisi}, {Anderson}, {Schneider}, {Simon}, {Kabanovic}, {G{\"u}sten}, {Zavagno}, {Broos}, {Buchbender}, {Guevara}, {Jacobs}, {Justen}, {Klein}, {Linville}, {R{\"o}llig}, {Russeil}, {Stutzki}, {Tiwari}, {Townsley}, \& {Tielens}}]{Luisi2021}
{Luisi}, M., {Anderson}, L.~D., {Schneider}, N., {et~al.} 2021, Science Advances, 7, eabe9511, \dodoi{10.1126/sciadv.abe9511}

\bibitem[{{Madden} {et~al.}(2006){Madden}, {Galliano}, {Jones}, \& {Sauvage}}]{Madden2006}
{Madden}, S.~C., {Galliano}, F., {Jones}, A.~P., \& {Sauvage}, M. 2006, \aap, 446, 877, \dodoi{10.1051/0004-6361:20053890}

\bibitem[{{Madden} {et~al.}(2020){Madden}, {Cormier}, {Hony}, {Lebouteiller}, {Abel}, {Galametz}, {De Looze}, {Chevance}, {Polles}, {Lee}, {Galliano}, {Lambert-Huyghe}, {Hu}, \& {Ramambason}}]{Madden2020}
{Madden}, S.~C., {Cormier}, D., {Hony}, S., {et~al.} 2020, \aap, 643, A141, \dodoi{10.1051/0004-6361/202038860}

\bibitem[{{Massey}(2002)}]{Massey2002}
{Massey}, P. 2002, \apjs, 141, 81, \dodoi{10.1086/338286}

\bibitem[{{Meixner} {et~al.}(2006){Meixner}, {Gordon}, {Indebetouw}, {Hora}, {Whitney}, {Blum}, {Reach}, {Bernard}, {Meade}, {Babler}, {Engelbracht}, {For}, {Misselt}, {Vijh}, {Leitherer}, {Cohen}, {Churchwell}, {Boulanger}, {Frogel}, {Fukui}, {Gallagher}, {Gorjian}, {Harris}, {Kelly}, {Kawamura}, {Kim}, {Latter}, {Madden}, {Markwick-Kemper}, {Mizuno}, {Mizuno}, {Mould}, {Nota}, {Oey}, {Olsen}, {Onishi}, {Paladini}, {Panagia}, {Perez-Gonzalez}, {Shibai}, {Sato}, {Smith}, {Staveley-Smith}, {Tielens}, {Ueta}, {van Dyk}, {Volk}, {Werner}, \& {Zaritsky}}]{Meixner2006}
{Meixner}, M., {Gordon}, K.~D., {Indebetouw}, R., {et~al.} 2006, \aj, 132, 2268, \dodoi{10.1086/508185}

\bibitem[{{Muller} {et~al.}(2003){Muller}, {Staveley-Smith}, {Zealey}, \& {Stanimirovi{\'c}}}]{Muller2003}
{Muller}, E., {Staveley-Smith}, L., {Zealey}, W., \& {Stanimirovi{\'c}}, S. 2003, \mnras, 339, 105, \dodoi{10.1046/j.1365-8711.2003.06147.x}

\bibitem[{{Nigra} {et~al.}(2008){Nigra}, {Gallagher}, {Smith}, {Stanimirovi{\'c}}, {Nota}, \& {Sabbi}}]{Nigra2008}
{Nigra}, L., {Gallagher}, J.~S., {Smith}, L.~J., {et~al.} 2008, \pasp, 120, 972, \dodoi{10.1086/592236}

\bibitem[{{Oliveira} {et~al.}(2013){Oliveira}, {van Loon}, {Sloan}, {Sewi{\l}o}, {Kraemer}, {Wood}, {Indebetouw}, {Filipovi{\'c}}, {Crawford}, {Wong}, {Hora}, {Meixner}, {Robitaille}, {Shiao}, \& {Simon}}]{Oliveira2013}
{Oliveira}, J.~M., {van Loon}, J.~T., {Sloan}, G.~C., {et~al.} 2013, \mnras, 428, 3001, \dodoi{10.1093/mnras/sts250}

\bibitem[{{Oliveira} {et~al.}(2019){Oliveira}, {van Loon}, {Sewi{\l}o}, {Lee}, {Lebouteiller}, {Chen}, {Cormier}, {Filipovi{\'c}}, {Carlson}, {Indebetouw}, {Madden}, {Meixner}, {Sargent}, \& {Fukui}}]{Oliveira2019}
{Oliveira}, J.~M., {van Loon}, J.~T., {Sewi{\l}o}, M., {et~al.} 2019, \mnras, 490, 3909, \dodoi{10.1093/mnras/stz2810}

\bibitem[{{O'Neill} {et~al.}(2022){O'Neill}, {Indebetouw}, {Sandstrom}, {Bolatto}, {Jameson}, {Carlson}, {Finn}, {Meixner}, {Sabbi}, \& {Sewi{\l}o}}]{ONeill2022}
{O'Neill}, T.~J., {Indebetouw}, R., {Sandstrom}, K., {et~al.} 2022, \apj, 938, 82, \dodoi{10.3847/1538-4357/ac8d93}

\bibitem[{{Perrin} {et~al.}(2015){Perrin}, {Long}, {Sivaramakrishnan}, {Lajoie}, {Elliot}, {Pueyo}, \& {Albert}}]{Perrin2015}
{Perrin}, M.~D., {Long}, J., {Sivaramakrishnan}, A., {et~al.} 2015, {WebbPSF: James Webb Space Telescope PSF Simulation Tool}, Astrophysics Source Code Library, record ascl:1504.007

\bibitem[{{Price-Whelan} {et~al.}(2018){Price-Whelan}, {Sip{\H{o}}cz}, {G{\"u}nther}, {Lim}, {Crawford}, {Conseil}, {Shupe}, {Craig}, {Dencheva}, {Ginsburg}, {VanderPlas}, {Bradley}, {P{\'e}rez-Su{\'a}rez}, {de Val-Borro}, {Paper Contributors}, {Aldcroft}, {Cruz}, {Robitaille}, {Tollerud}, {Coordination Committee}, {Ardelean}, {Babej}, {Bach}, {Bachetti}, {Bakanov}, {Bamford}, {Barentsen}, {Barmby}, {Baumbach}, {Berry}, {Biscani}, {Boquien}, {Bostroem}, {Bouma}, {Brammer}, {Bray}, {Breytenbach}, {Buddelmeijer}, {Burke}, {Calderone}, {Cano Rodr{\'\i}guez}, {Cara}, {Cardoso}, {Cheedella}, {Copin}, {Corrales}, {Crichton}, {D{\textquoteright}Avella}, {Deil}, {Depagne}, {Dietrich}, {Donath}, {Droettboom}, {Earl}, {Erben}, {Fabbro}, {Ferreira}, {Finethy}, {Fox}, {Garrison}, {Gibbons}, {Goldstein}, {Gommers}, {Greco}, {Greenfield}, {Groener}, {Grollier}, {Hagen}, {Hirst}, {Homeier}, {Horton}, {Hosseinzadeh}, {Hu}, {Hunkeler}, {Ivezi{\'c}}, {Jain}, {Jenness}, {Kanarek}, {Kendrew}, {Kern}, {Kerzendorf}, {Khvalko},
  {King}, {Kirkby}, {Kulkarni}, {Kumar}, {Lee}, {Lenz}, {Littlefair}, {Ma}, {Macleod}, {Mastropietro}, {McCully}, {Montagnac}, {Morris}, {Mueller}, {Mumford}, {Muna}, {Murphy}, {Nelson}, {Nguyen}, {Ninan}, {N{\"o}the}, {Ogaz}, {Oh}, {Parejko}, {Parley}, {Pascual}, {Patil}, {Patil}, {Plunkett}, {Prochaska}, {Rastogi}, {Reddy Janga}, {Sabater}, {Sakurikar}, {Seifert}, {Sherbert}, {Sherwood-Taylor}, {Shih}, {Sick}, {Silbiger}, {Singanamalla}, {Singer}, {Sladen}, {Sooley}, {Sornarajah}, {Streicher}, {Teuben}, {Thomas}, {Tremblay}, {Turner}, {Terr{\'o}n}, {van Kerkwijk}, {de la Vega}, {Watkins}, {Weaver}, {Whitmore}, {Woillez}, {Zabalza}, \& {Contributors}}]{astropy:2018}
{Price-Whelan}, A.~M., {Sip{\H{o}}cz}, B.~M., {G{\"u}nther}, H.~M., {et~al.} 2018, \aj, 156, 123, \dodoi{10.3847/1538-3881/aabc4f}

\bibitem[{{Ramachandran} {et~al.}(2019){Ramachandran}, {Hamann}, {Oskinova}, {Gallagher}, {Hainich}, {Shenar}, {Sander}, {Todt}, \& {Fulmer}}]{Ramachandran2019}
{Ramachandran}, V., {Hamann}, W.~R., {Oskinova}, L.~M., {et~al.} 2019, \aap, 625, A104, \dodoi{10.1051/0004-6361/201935365}

\bibitem[{Reiter \& Parker(2019)}]{Reiter2019}
Reiter, M., \& Parker, R.~J. 2019, Monthly Notices of the Royal Astronomical Society, 486, 4354, \dodoi{10.1093/mnras/stz1115}

\bibitem[{{Richardson} {et~al.}(2024){Richardson}, {Ginsburg}, {Indebetouw}, \& {Robitaille}}]{Richardson2024}
{Richardson}, T., {Ginsburg}, A., {Indebetouw}, R., \& {Robitaille}, T.~P. 2024, \apj, 961, 188, \dodoi{10.3847/1538-4357/ad072d}

\bibitem[{Robitaille(2017)}]{Robitaille2017}
Robitaille, T.~P. 2017, Astronomy and Astrophysics, 600, \dodoi{10.1051/0004-6361/201425486}

\bibitem[{{Robitaille} {et~al.}(2006){Robitaille}, {Whitney}, {Indebetouw}, {Wood}, \& {Denzmore}}]{Robitaille2006}
{Robitaille}, T.~P., {Whitney}, B.~A., {Indebetouw}, R., {Wood}, K., \& {Denzmore}, P. 2006, \apjs, 167, 256, \dodoi{10.1086/508424}

\bibitem[{Roman-Duval {et~al.}(2014)Roman-Duval, Gordon, Meixner, Bot, Bolatto, Hughes, Wong, Babler, Bernard, Clayton, Fukui, Galametz, Galliano, Glover, Hony, Israel, Jameson, Lebouteiller, Lee, Li, Madden, Misselt, Montiel, Okumura, Onishi, Panuzzo, Reach, Remy-Ruyer, Robitaille, Rubio, Sauvage, Seale, Sewilo, Staveley-Smith, \& Zhukovska}]{RomanDuval2014}
Roman-Duval, J., Gordon, K.~D., Meixner, M., {et~al.} 2014, The Astrophysical Journal, 797, 86, \dodoi{10.1088/0004-637X/797/2/86}

\bibitem[{{Russell} \& {Dopita}(1992)}]{Russell1992}
{Russell}, S.~C., \& {Dopita}, M.~A. 1992, \apj, 384, 508, \dodoi{10.1086/170893}

\bibitem[{{Salpeter}(1955)}]{Salpeter1955}
{Salpeter}, E.~E. 1955, \apj, 121, 161, \dodoi{10.1086/145971}

\bibitem[{{Sanduleak}(1969)}]{Sanduleak1969}
{Sanduleak}, N. 1969, \aj, 74, 877, \dodoi{10.1086/110875}

\bibitem[{{Schmalzl} {et~al.}(2008){Schmalzl}, {Gouliermis}, {Dolphin}, \& {Henning}}]{Schmalzl2008}
{Schmalzl}, M., {Gouliermis}, D.~A., {Dolphin}, A.~E., \& {Henning}, T. 2008, \apj, 681, 290, \dodoi{10.1086/588722}

\bibitem[{{Seale} {et~al.}(2014){Seale}, {Meixner}, {Sewi{\l}o}, {Babler}, {Engelbracht}, {Gordon}, {Hony}, {Misselt}, {Montiel}, {Okumura}, {Panuzzo}, {Roman-Duval}, {Sauvage}, {Boyer}, {Chen}, {Indebetouw}, {Matsuura}, {Oliveira}, {Srinivasan}, {van Loon}, {Whitney}, \& {Woods}}]{Seale2014}
{Seale}, J.~P., {Meixner}, M., {Sewi{\l}o}, M., {et~al.} 2014, \aj, 148, 124, \dodoi{10.1088/0004-6256/148/6/124}

\bibitem[{{Selier} {et~al.}(2011){Selier}, {Heydari-Malayeri}, \& {Gouliermis}}]{Selier2011}
{Selier}, R., {Heydari-Malayeri}, M., \& {Gouliermis}, D.~A. 2011, \aap, 529, A40, \dodoi{10.1051/0004-6361/201016100}

\bibitem[{{Sewi{\l}o} {et~al.}(2013){Sewi{\l}o}, {Carlson}, {Seale}, {Indebetouw}, {Meixner}, {Whitney}, {Robitaille}, {Oliveira}, {Gordon}, {Meade}, {Babler}, {Hora}, {Block}, {Misselt}, {van Loon}, {Chen}, {Churchwell}, \& {Shiao}}]{Sewilo2013}
{Sewi{\l}o}, M., {Carlson}, L.~R., {Seale}, J.~P., {et~al.} 2013, \apj, 778, 15, \dodoi{10.1088/0004-637X/778/1/15}

\bibitem[{{Smithsonian Astrophysical Observatory}(2000)}]{ds92000}
{Smithsonian Astrophysical Observatory}. 2000, {SAOImage DS9: A utility for displaying astronomical images in the X11 window environment}.
\newblock \doeprint{0003.002}

\bibitem[{{Stanimirovic} {et~al.}(2000){Stanimirovic}, {Staveley-Smith}, {van der Hulst}, {Bontekoe}, {Kester}, \& {Jones}}]{Stanimirovic2000}
{Stanimirovic}, S., {Staveley-Smith}, L., {van der Hulst}, J.~M., {et~al.} 2000, \mnras, 315, 791, \dodoi{10.1046/j.1365-8711.2000.03480.x}

\bibitem[{{Stephens} {et~al.}(2017){Stephens}, {Gouliermis}, {Looney}, {Gruendl}, {Chu}, {Weisz}, {Seale}, {Chen}, {Wong}, {Hughes}, {Pineda}, {Ott}, \& {Muller}}]{Stephens2017}
{Stephens}, I.~W., {Gouliermis}, D., {Looney}, L.~W., {et~al.} 2017, \apj, 834, 94, \dodoi{10.3847/1538-4357/834/1/94}

\bibitem[{{Stern} {et~al.}(2005){Stern}, {Eisenhardt}, {Gorjian}, {Kochanek}, {Caldwell}, {Eisenstein}, {Brodwin}, {Brown}, {Cool}, {Dey}, {Green}, {Jannuzi}, {Murray}, {Pahre}, \& {Willner}}]{Stern2005}
{Stern}, D., {Eisenhardt}, P., {Gorjian}, V., {et~al.} 2005, \apj, 631, 163, \dodoi{10.1086/432523}

\bibitem[{{Testor} {et~al.}(2014){Testor}, {Heydari-Malayeri}, {Chen}, {Lemaire}, {Sewi{\l}o}, \& {Diana}}]{Testor2014}
{Testor}, G., {Heydari-Malayeri}, M., {Chen}, C. H.~R., {et~al.} 2014, \aap, 564, A31, \dodoi{10.1051/0004-6361/201118484}

\bibitem[{{van der Wel} {et~al.}(2014){van der Wel}, {Franx}, {van Dokkum}, {Skelton}, {Momcheva}, {Whitaker}, {Brammer}, {Bell}, {Rix}, {Wuyts}, {Ferguson}, {Holden}, {Barro}, {Koekemoer}, {Chang}, {McGrath}, {H{\"a}ussler}, {Dekel}, {Behroozi}, {Fumagalli}, {Leja}, {Lundgren}, {Maseda}, {Nelson}, {Wake}, {Patel}, {Labb{\'e}}, {Faber}, {Grogin}, \& {Kocevski}}]{vanderWel2014}
{van der Wel}, A., {Franx}, M., {van Dokkum}, P.~G., {et~al.} 2014, \apj, 788, 28, \dodoi{10.1088/0004-637X/788/1/28}

\bibitem[{Van~Rossum \& Drake(2009)}]{VanRossum2009}
Van~Rossum, G., \& Drake, F.~L. 2009, Python 3 Reference Manual (Scotts Valley, CA: CreateSpace)

\bibitem[{{Ward} {et~al.}(2017){Ward}, {Oliveira}, {van Loon}, \& {Sewi{\l}o}}]{Ward2017}
{Ward}, J.~L., {Oliveira}, J.~M., {van Loon}, J.~T., \& {Sewi{\l}o}, M. 2017, \mnras, 464, 1512, \dodoi{10.1093/mnras/stw2386}

\bibitem[{{Wolfire} {et~al.}(2010){Wolfire}, {Hollenbach}, \& {McKee}}]{Wolfire2010}
{Wolfire}, M.~G., {Hollenbach}, D., \& {McKee}, C.~F. 2010, \apj, 716, 1191, \dodoi{10.1088/0004-637X/716/2/1191}

\bibitem[{{Zeidler} {et~al.}(2024){Zeidler}, {Sabbi}, {Nota}, {Manjavacas}, {Jones}, \& {Pacifici}}]{Zeidler2024}
{Zeidler}, P., {Sabbi}, E., {Nota}, A., {et~al.} 2024, \apj, 975, 18, \dodoi{10.3847/1538-4357/ad779e}

\end{thebibliography}
\bibliographystyle{aasjournal}

\appendix
\restartappendixnumbering

\section{YSO/SED models} \label{app:ysomodel}

\begin{figure}[ht!]
\centering
\subfigure[MZS-1b]{
\includegraphics[width=0.4\textwidth]{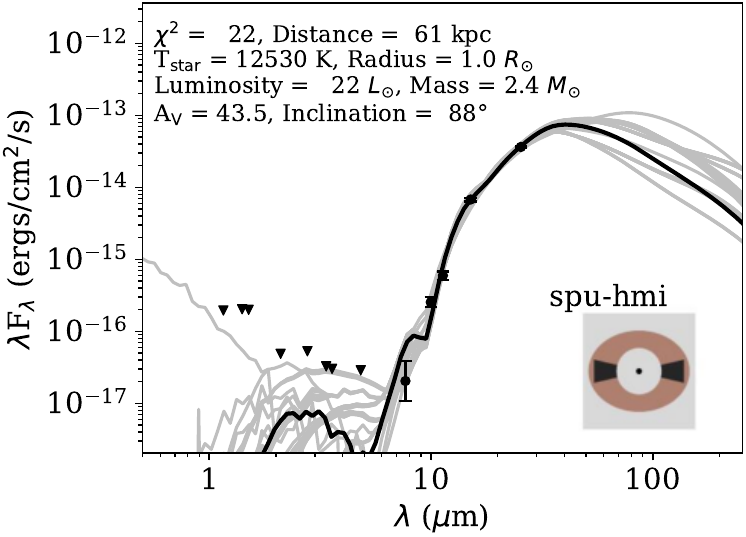}}
\subfigure[MZS-1c]{
\includegraphics[width=0.4\textwidth]{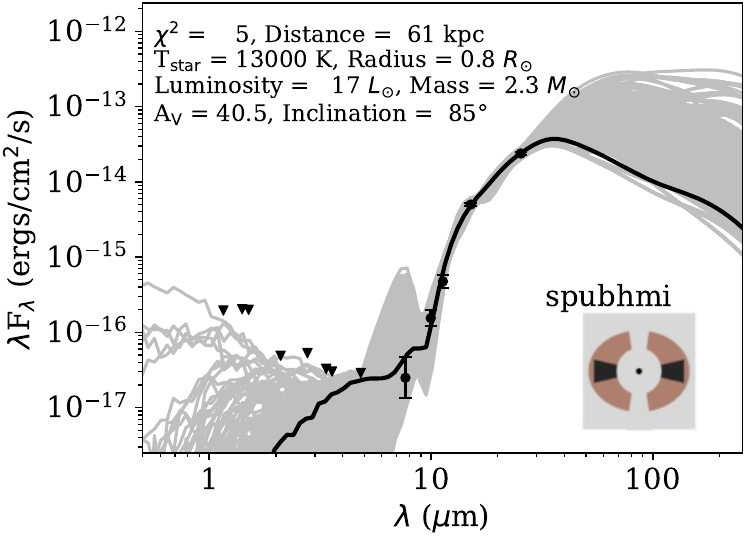}}

\subfigure[MZS-1d]{
\includegraphics[width=0.4\textwidth]{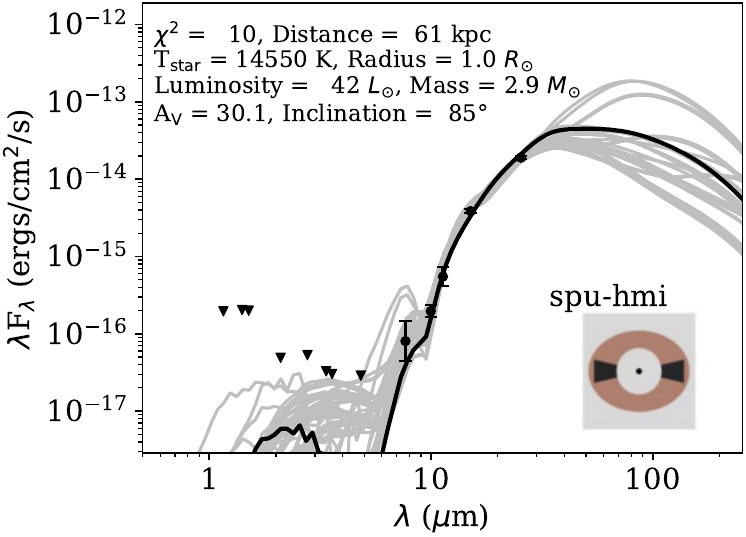}}
\subfigure[MZS-1e]{
\includegraphics[width=0.4\textwidth]{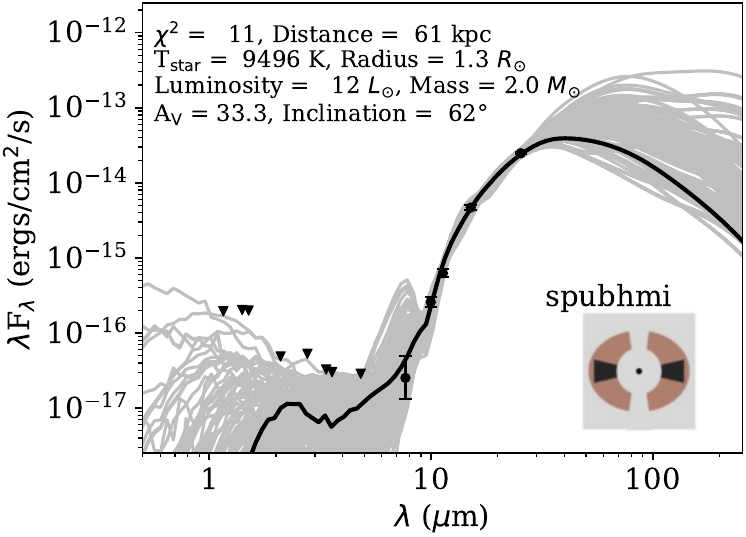}
}
\subfigure[MZS-1f]{
\includegraphics[width=0.4\textwidth]{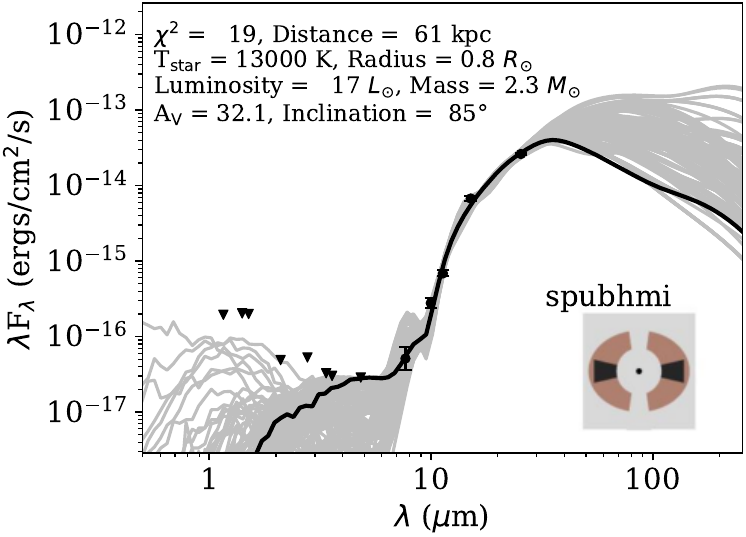}}
\caption{Same as Figure~\ref{fig:sed_fits}, but for the remaining candidate YSOs in MZS-1. Each panel shows SED fits using models from \citet{Robitaille2017}, based on the simplest and most likely model set as indicated in the inset. The black curve represents the best-fit model determined by $\chi^2$ minimization, with the corresponding parameters listed in the top-left corner. The gray curves show the other models within the same grid, which are used to estimate the parameter ranges reported in Table~\ref{tab1:YSO_params}.} 
\label{fig:app1_yso}
\end{figure}

Figures~\ref{fig:app1_yso} present the most likely model fits for the YSO candidates MZS-1b, MZS-1c, MZS-1d, MZS-1e, and MZS-1f, as discussed in Section~\ref{subsec:yso_fit}. These fits are used to derive the ranges of stellar parameters—temperatures, luminosities, and radii—reported in Table~\ref{tab1:YSO_params}.

\end{document}